\documentclass[journal]{IEEEtran}
\usepackage{amsmath}
\usepackage{graphicx}
\usepackage{epsfig}
\usepackage{amssymb}
\usepackage{amsthm}
\usepackage{verbatim}
\usepackage{lettrine}
\usepackage{flushend,cuted}
\usepackage[top=1in, bottom=0.8in, left=0.5in, right=0.5in]{geometry}

\usepackage{url}
\usepackage{multirow}
\usepackage{cite}
\usepackage{mathrsfs}
\usepackage{cases}
\usepackage{caption}
\usepackage[english]{babel}
\usepackage[utf8]{inputenc}
\usepackage{tikz}
\usepackage{nomencl}
\makenomenclature

\begin{document}

\title{Transient Stability Assessment Using Individual Machine Equal Area Criterion Part I: Unity Principle
}
{\author{Songyan Wang, Jilai Yu, Wei Zhang \emph{ Member, IEEE}

\thanks{

S. Wang is with Department of Electrical Engineering, Harbin Institute of Technology, Harbin 150001, China (e-mail: wangsongyan@163.com).

J. Yu is with Department of Electrical Engineering, Harbin Institute of Technology, Harbin 150001, China (e-mail: yupwrs@hit.edu.cn).

W. Zhang is with Department of Electrical Engineering, Harbin Institute of Technology, Harbin 150001, China (e-mail: wzps@hit.edu.cn).
}
}
\maketitle
\vspace{-15pt}
\begin{abstract}

Analyzing system trajectory from the perspective of individual machines provides a distinctive angle to analyze the transient stability of power systems. This two-paper series propose a direct-time-domain method that is based on the individual-machine equal area criterion. In the first paper, by examining the mapping between the trajectory and power-vs-angle curve of an individual machine, the stability property to characterize a critical machine is clarified. The mapping between the system trajectory and individual-machine equal area criterion is established. Furthermore, a unity principle between the individual-machine stability and system stability is proposed. It is proved that the instability of the system can be confirmed by finding any one unstable critical machine, thence, the transient stability of a multimachine system can be monitored in an individual-machine way in transient stability assessment.

\end{abstract}


\nomenclature[1]{IMEF}{Individual machine energy function}
\nomenclature[2]{PEF}	{Partial energy function}
\nomenclature[3]{OMIB}	{One-machine-infinite-bus}
\nomenclature[4]{EAC}	{Equal area criterion}
\nomenclature[5]{COI}	{Center of inertia}
\nomenclature[6]{DLP}	{Dynamic liberation point}
\nomenclature[7]{DSP}	{Dynamic stationary point}
\nomenclature[8]{UEP}	{Unstable equilibrium point}
\nomenclature[9]{RUEP}	{Relevant UEP}
\nomenclature[10]{CUEP}	{Controlling UEP}
\nomenclature[11]{SVCS}	{Single-machine-and-virtual-COI-machine subsystem}
\nomenclature[12]{EEAC}	{Extended equal area criterion}
\nomenclature[13]{SIME}	{Single machine equivalence}
\nomenclature[14]{MOD}	{Mode of disturbance}
\nomenclature[15]{TSA}	{Transient stability assessment}
\nomenclature[16]{P.E.}	{Potential energy}
\nomenclature[17]{K.E.}	{Kinetic energy}
\nomenclature[18]{CCT}	{Critical fault clearing time}

\printnomenclature

\begin{center}
  \textsc{Abbreviation}
\end{center}
\begin{table}[!htbp]
\captionsetup{name=\textsc{Table}}
\normalsize
\vspace{5pt}
\setlength{\belowcaptionskip}{0pt}
\vspace{-4pt}
\setlength{\abovecaptionskip}{2pt}
\begin{tabular}{ll}
{COI} & {Center of inertia} \\
{CCT}&{Critical clearing time} \\
{CDSP}&{DSP of the critical stable machine}\\
{CUEP}&{Controlling UEP}\\
{DLP}&{Dynamic liberation point}\\
{DSP}&{Dynamic stationary point}\\
{EAC}&{Equal area criterion}\\
{IEEAC}&{Integrated extended EAC}\\
{IMEAC}&{Individual-machine EAC}\\
{IMEF}&{Individual machine energy function}\\
{IVCS}&{Individual machine-virtual COI machine system}\\
{LOSP}&{Loss-of-synchronism point}\\
{OMIB}&{One-machine-infinite-bus}\\
{PEF}&{Partial energy function}\\
{TSA}&{Transient stability assessment}\\
{UEP}&{Unstable equilibrium point}\\
\end{tabular}
\end{table}

\section{Introduction}

\subsection{Literature Review}
In the last decades, great efforts have been made to the application of direct methods in the transient stability analysis. In early work, Lyapunov method was proved to yield conservative results \cite{Fouad1975Stability}. Hereafter, direct methods such as CUEP method, sustained fault method and IEEAC method had received considerable attention and achieved advances \cite{Athay1979A,Kakimoto1978Transient,Xue1989Extended,Jakpattanajit2016Enhancing}. These methods monitor transient behaviors of all machines in the system in TSA, which are also named as global methods.
Unlike global methods, some transient stability analysts incline to observe the power system transient stability from a distinctive individual-machine angle as “the instability of a multi-machine system is determined by the motion of some unstable critical machines if more than one machine tends to lose synchronism” \cite{Fouad1981Transient}. Stimulated by this concept of individual-machine perspective, Vittal and Fouad \cite{Vittal1982Power,Michel1982Power} stated that the instability of the system depends on the transient energy of individual machines and IMEF was proposed. Stanton \cite{Stanton1982Assessment,Stanton1989Analysis} performed a detailed machine-by-machine analysis of a multi-machine instability, and PEF was used to quantify the energy of a local control action. Later, EAC of the critical machine is applied in PEF method to identify system stability \cite{Stanton1995Application}. Rastgoufard et al. \cite{Rastgoufard1988Multi} used an IMEF in synchronous reference to determine the transient stability of a multimachine system. Haque \cite{Haque1995Further} proposed an efficient individual-machine method to compute the CCT of the system under transients. Ando and Iwamoto [15] presented a potential energy ridge which can be used to predict the single-machine stability.
Among all these works of individual-machine methods, the PEF method is quite representative and can be seen as a milestone because Ref. \cite{Stanton1989Analysis,Stanton1989Transient} initially explained fundamental theories and also provided some valuable underlying hypothesis regarding the individual-machine methods. Although individual-machine methods were proved to be effective for the stability analysis, these methods were at a standstill for decades. The reason is that some crucial concepts of the individual-machine methods were missing or were illustrated in an unsystematic and tutorial form, leaving confusions unclarified and controversial problems unsolved when they are used for TSA.

\subsection{Scope and contribution of the paper }

 In this two-paper series a direct-time-domain method that is based on individual-machine equal area criterion (IMEAC) is proposed. The first paper systematically clarifies the mechanism of the proposed method to monitor the transient stability of a multi-machine system. The companion paper applies the proposed method for TSA and CCT computation. In this paper, based on the actual trajectory of the system trajectory during transients, the Kimbark curve of a critical machine is first analyzed, and then IMEAC is proved to strictly hold for a critical machine. Second, following trajectory stability theory the concept of individual-machine trajectory (IMT) is proposed, and the mapping between system trajectory and Kimbark curve of an individual machine is established. In the end of the paper, the unity principle between individual-machine stability and system stability is proposed. It shows that IMT of any one unstable critical machine can drive system trajectory to go unstable, thence, and the transient stability of a multi-machine system can be monitored in an individual-machine way during TSA.
Contributions of this paper are summarized as follows:

(i) Following trajectory stability theory, this paper explains the mechanism of the transient stability of a multi-machine system from an individual-machine angle. The transient instability of the system can be determined by the instability of any one unstable critical machine;

(ii) Some mistakes about stability-characterization of a critical machine in Ref. \cite{Stanton1989Analysis,Stanton1989Transient} are corrected in this paper;

(iii) The unity principle explicitly explains the relationship between individual-machine stability and system stability is proposed. This may release the potential of the usage of individual-machine monitoring in TSA.

In this paper we only discuss the first-swing stability of a critical machine, and swing stability in this paper only depicts the stability state of a critical machine when the velocity of the machine reaches zero, rather than following the conventional global concept.
Three test systems are applied in this two-paper series. Test System-1 (TS-1) is a modified IEEE 39-bus system. In TS-1 the inertia constant of Unit 39 is modified to 200 p.u. from 1000 p.u.; Test System-2 (TS-2) is the standard IEEE 118-bus system. Test System-3 (TS-3) is a practical 2766-bus interconnected system. All faults are three phase short-circuits faults which occurred at 0 s and they are cleared without line switching. Fault types in this two-paper series are described in the form of [test-system, fault location, fault-on time]. The simulations of TS-1 and that of TS-2 are fully based on the classical model given in \cite{Jakpattanajit2016Enhancing}. The simulations of TS-3 are based on complicated dynamic models, and the parameters of this system can be found in the companion paper.

The remaining paper is organized as follows. In Section II, IMEAC of a critical machine is analyzed. In Section III, the Kimbark curves of both critical machines and non-critical machines that rely on actual system trajectory in the multi-machine system are analyzed. In Section IV, the mapping between IMT and the Kimbark curve of a critical machine is established. In Section V, the unity principle of system stability and stability of a critical machine is depicted in the sense of IMT. In Section VI, an example about the application of the proposed method is demonstrated. Conclusions are provided in Section VII.

\section{Individual-Machine Equal Area Criterion}

\subsection{Equation of Motion of an individual machine}

Conventionally, for a direct method that is based on COI reference, ``an individual machine” should be precisely expressed as ``an individual machine in COI reference”. For a n-machine system with rotor angle $\delta_{i}$ and inertia constant $M_i$, the motion of an individual machine $i$ in the synchronous reference is governed by differential equations:

\begin{equation}\label{Eq_Mac_Mot}
   \begin{cases}
   \dot{\delta_{i}}=\omega_{i} \\
    M_{i}\dot{\omega_{i}}=P_{mi}-P_{ei}\\
   \end{cases}
     \setlength{\abovedisplayskip}{1pt}
  \setlength{\belowdisplayskip}{1pt}
\end{equation}

Position of the COI of the system is defined by:

\begin{equation}\label{Eq_Pos_COI}
   \begin{cases}
   \dot{\delta}_{COI}=\frac{1}{M_{T}}\sum\limits_{i=1}^{n}M_{i}\delta_{i} \\
   \omega_{COI}= \frac{1}{M_{T}}\sum\limits_{i=1}^{n}M_{i}\omega_{i} \\
    P_{COI}= \sum\limits_{i=1}^{n}(P_{mi}-P_{ei}) \\
   \end{cases}
     \setlength{\abovedisplayskip}{1pt}
  \setlength{\belowdisplayskip}{1pt}
\end{equation}
where $M_{T}=\sum\limits_{i=1}^{n}M_{i}$.

From (\ref{Eq_Pos_COI}), the motion of COI is determined by:

\begin{equation}\label{Eq_Mot_COI}
   \begin{cases}
   \dot{\delta}_{COI}=\omega_{COI} \\
    M_{T}\dot{\omega}_{COI}=P_{COI}\\
   \end{cases}
     \setlength{\abovedisplayskip}{1pt}
  \setlength{\belowdisplayskip}{1pt}
\end{equation}

Eqn. (\ref{Eq_Mot_COI}) indicates that COI can also be seen as a virtual ``machine” with its own equation of motion being described as the aggregated motion of all machines in the system.The trajectory of the virtual COI machine in synchronous reference is shown in Fig. 1.

Following (\ref{Eq_Mac_Mot}) and (\ref{Eq_Mot_COI}), since machine $i$ and COI are two ``single” machines with interactions, a two-machine subsystem can be formed by using these two machines, which is defined as a SVCS, as shown in Fig. \ref{Fig_Tra_Vir_Mac_1}.

\begin{figure}
\vspace{5pt}
\captionsetup{name=Fig.,font={small},singlelinecheck=off,justification=raggedright}
\includegraphics[width=3.5in,height=2.8in,keepaspectratio]{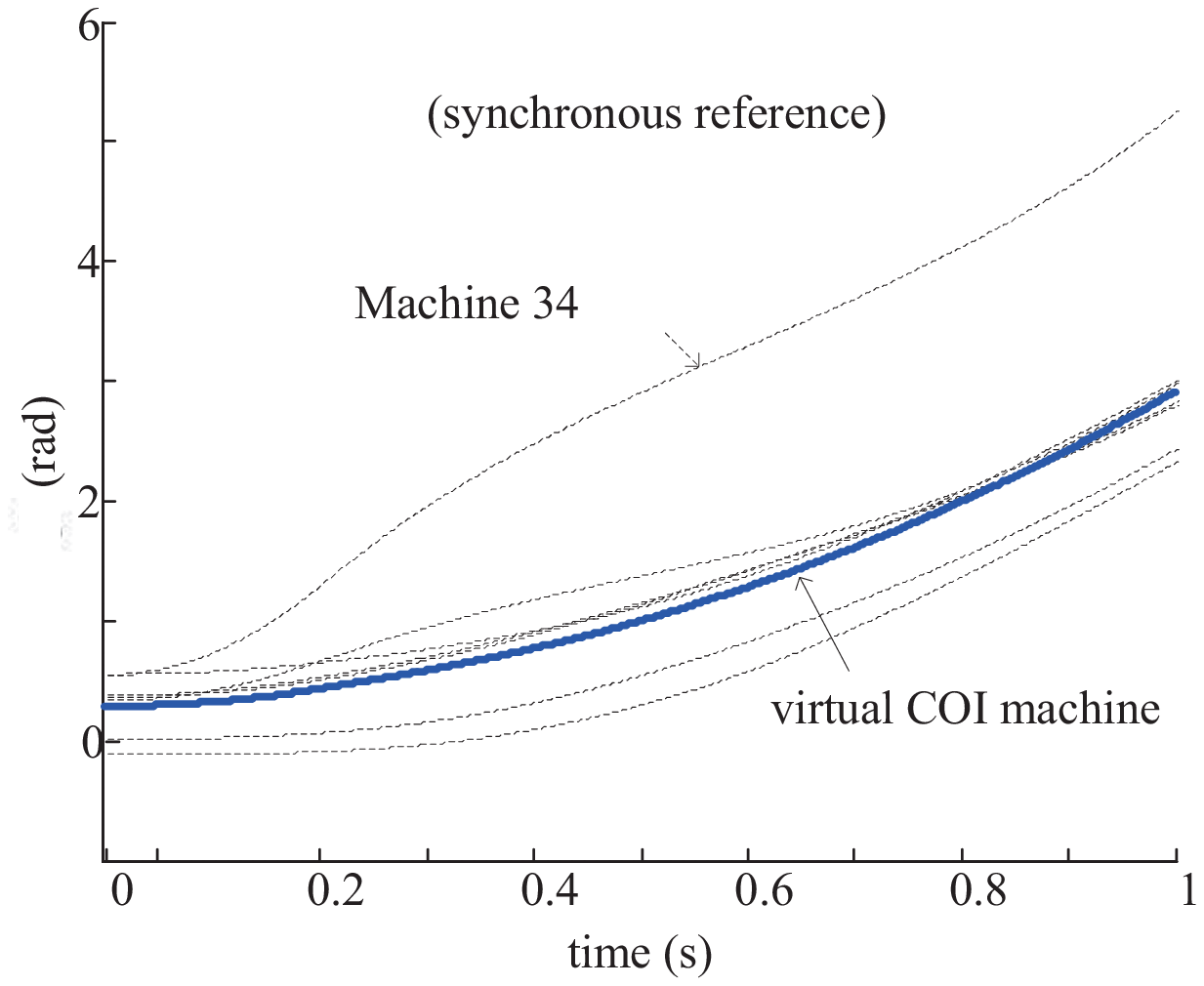}\\
  \setlength{\abovecaptionskip}{-5pt}
  \setlength{\belowcaptionskip}{0pt}
  \vspace{-2pt}
  \caption{ Trajectory of the virtual COI machine [TS-1, bus-34, 0.202s]}
  \label{Fig_Tra_Vir_Mac_1}
\end{figure}

Since machine $i$ and COI are two ``individual” machines with interactions, a two-machine system which is named as Individual machine-virtual COI machine system (IVCS) can be formed by these two machines, as in Fig. \ref{Fig_Two_Mac_COI_2}. The relative trajectory between a critical machine and the virtual COI machine in an IVCS is shown in Fig. \ref{Fig_Ins_Cri_Vir_3}.

\begin{figure}
\vspace{5pt}
\captionsetup{name=Fig.,font={small},singlelinecheck=off,justification=raggedright}
\includegraphics[width=3.5in,height=2.8in,keepaspectratio]{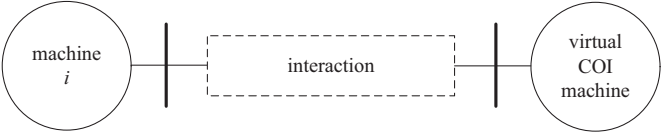}\\
  \setlength{\abovecaptionskip}{-5pt}
  \setlength{\belowcaptionskip}{0pt}
  \vspace{-2pt}
  \caption{ Two-machine system formed by machine $i$ and virtual COI machine.}
  \label{Fig_Two_Mac_COI_2}
\end{figure}

\begin{figure}
\vspace{5pt}
\captionsetup{name=Fig.,font={small},singlelinecheck=off,justification=raggedright}
\includegraphics[width=3.5in,height=2.8in,keepaspectratio]{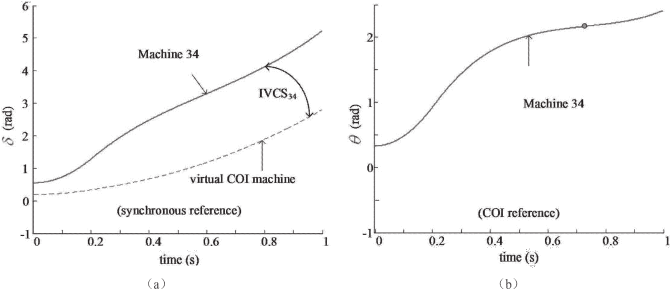}\\
  \setlength{\abovecaptionskip}{-5pt}
  \setlength{\belowcaptionskip}{0pt}
  \vspace{-2pt}
  \caption{ The instability of the relative trajectory between a critical machine and the virtual COI machine [TS-1, bus-34, 0.202s]. (a) an IVCS in synchronous reference. (b) an individual machine in COI reference}
  \label{Fig_Ins_Cri_Vir_3}
\end{figure}

Following (\ref{Eq_Mac_Mot}) and (\ref{Eq_Mot_COI}) , the relative motion between machine $i$ and the virtual COI machine of the SVCS can be given as:
\begin{equation}\label{Eq_Mot_SVC}
   \begin{cases}
   \dot{\theta}_{i}=\tilde{\omega}_{i} \\
    M_{i}\dot{\tilde{\omega}}_{i}=f_{i}\\
   \end{cases}
     \setlength{\abovedisplayskip}{1pt}
  \setlength{\belowdisplayskip}{1pt}
\end{equation}
where
\newline
$f_{i}=P_{mi}-P_{ei}-\frac{M_{i}}{M_{T}}P_{COI}$
\newline
$\theta{i}=\delta_{i}-\delta_{COI}$
\newline
$\tilde{\omega}_{i}=\omega_{i}-\omega_{COI}$

Eq. (\ref{Eq_Mot_SVC}) depicts the separation of an individual machine with respect to the COI of all machines in the system, which is substantially identical to the motion of an individual machine in COI reference. Therefore, each individual machine in COI reference should be precisely depicted as an IVCS that is formed by a ``pair” of machines. Yet, in this paper for historical reasons the ``individual machine” or ``critical machine” is still used in default for simplification.

\subsection{Strict EAC characteristic of an individual machine }

Since EAC only strictly holds in the OMIB system and the two-machine system, the above analysis given in Section II.A is of value because it indicates that EAC strictly holds for an individual machine as the IVCS is a precise two-machine system.

Following (\ref{Eq_Mot_SVC}), we have:

\begin{equation}\label{Eq_EAC_Dif}
f_{i}d\theta_{i}= M_{i}\tilde{\omega}_{i}d\tilde{\omega}_{i}
  \setlength{\abovedisplayskip}{1pt}
  \setlength{\belowdisplayskip}{1pt}
\end{equation}

Along the actual fault-on trajectory until fault clearing, we have

\begin{equation}\label{Eq_EAC_Int}
\int_{\theta_{i}^{0}}^{\theta_{i}^{c}}(f_{i}^{(F)})d\theta_{i}= \int_{\tilde{\omega}_{ij}^{0}}^{\tilde{\omega}_{i}^{c}}M_{i}\tilde{\omega}_{i}d\tilde{\omega}_{i}
  \setlength{\abovedisplayskip}{1pt}
  \setlength{\belowdisplayskip}{1pt}
\end{equation}
where $f_{i}^{(F)}$ corresponds to $f_{i}$ during fault-on period.

Eqn. (\ref{Eq_EAC_Int}) can be further expressed as:

\begin{equation}\label{Eq_EAC_Pos}
\frac{1}{2}M_{i}{\tilde{\omega_{i}^{c}}}^{2}=\int_{\theta_{i}^{0}}^{\theta_{ij}^{c}}\{0-(-f_{i}^{(F)})\}d\theta_{i}
  \setlength{\abovedisplayskip}{1pt}
  \setlength{\belowdisplayskip}{1pt}
\end{equation}

Along the actual post-fault trajectory after fault clearing, we have:

\begin{equation}\label{Eq_EAC_Pos_Act}
\int_{\theta_{i^{c}}}^{\theta_{i}}f_{i}^{(PF)}d\theta_{i}= \int_{\tilde{\omega_{i}^{c}}}^{\tilde{\omega_{i}}}M_{i}\tilde{\omega_{i}}d\tilde{\omega_{i}}
  \setlength{\abovedisplayskip}{1pt}
  \setlength{\belowdisplayskip}{1pt}
\end{equation}
where $f_{i}^{(PF)}$  corresponds to $f_{i}$ during post-fault period.

Eqn. (\ref{Eq_EAC_Pos_Act}) can be further expressed as:
\begin{equation}\label{Eq_EAC_Pos_1}
\frac{1}{2}M_{i}\tilde{\omega_{i}^{(c)}}^{2}=\int_{\theta_{i^{c}}}^{\theta_{i}}\{-f_{i}^{(PF)}-0\}d\theta_{i}+\frac{1}{2}M_{i}\tilde{\omega_{i}}^{2}
  \setlength{\abovedisplayskip}{1pt}
  \setlength{\belowdisplayskip}{1pt}
\end{equation}

Substituting (\ref{Eq_EAC_Pos}) into (\ref{Eq_EAC_Pos_1}) yields:
\begin{equation}\label{Eq_EAC_Fal}
\int_{\theta_{i}^{0}}^{\theta_{ij}^{c}}\{0-(-f_{i}^{(F)})\}d\theta_{i}=\frac{1}{2}M_{i}\tilde{\omega_{i}}^{2}+\int_{\theta_{i^{c}}}^{\theta_{i}}\{-f_{i}^{(PF)}-0\}d\theta_{i}
  \setlength{\abovedisplayskip}{1pt}
  \setlength{\belowdisplayskip}{1pt}
\end{equation}

A typical Kimbark curve \cite{Stanton1989Transient} (power-vs-angle curve) of an unstable critical machine that is formulated with the actual simulated system trajectory in the $\theta_{i}-f_{i}$ space is shown in Fig. \ref{Fig_Sim_Fal_4} (a). The rotor angles of the system are shown in Fig. \ref{Fig_Sim_Fal_4} (b).

\begin{figure}
\captionsetup{name=Fig.,font={small},singlelinecheck=off,justification=raggedright}
  \includegraphics[width=3.5in,height=5.4in,keepaspectratio]{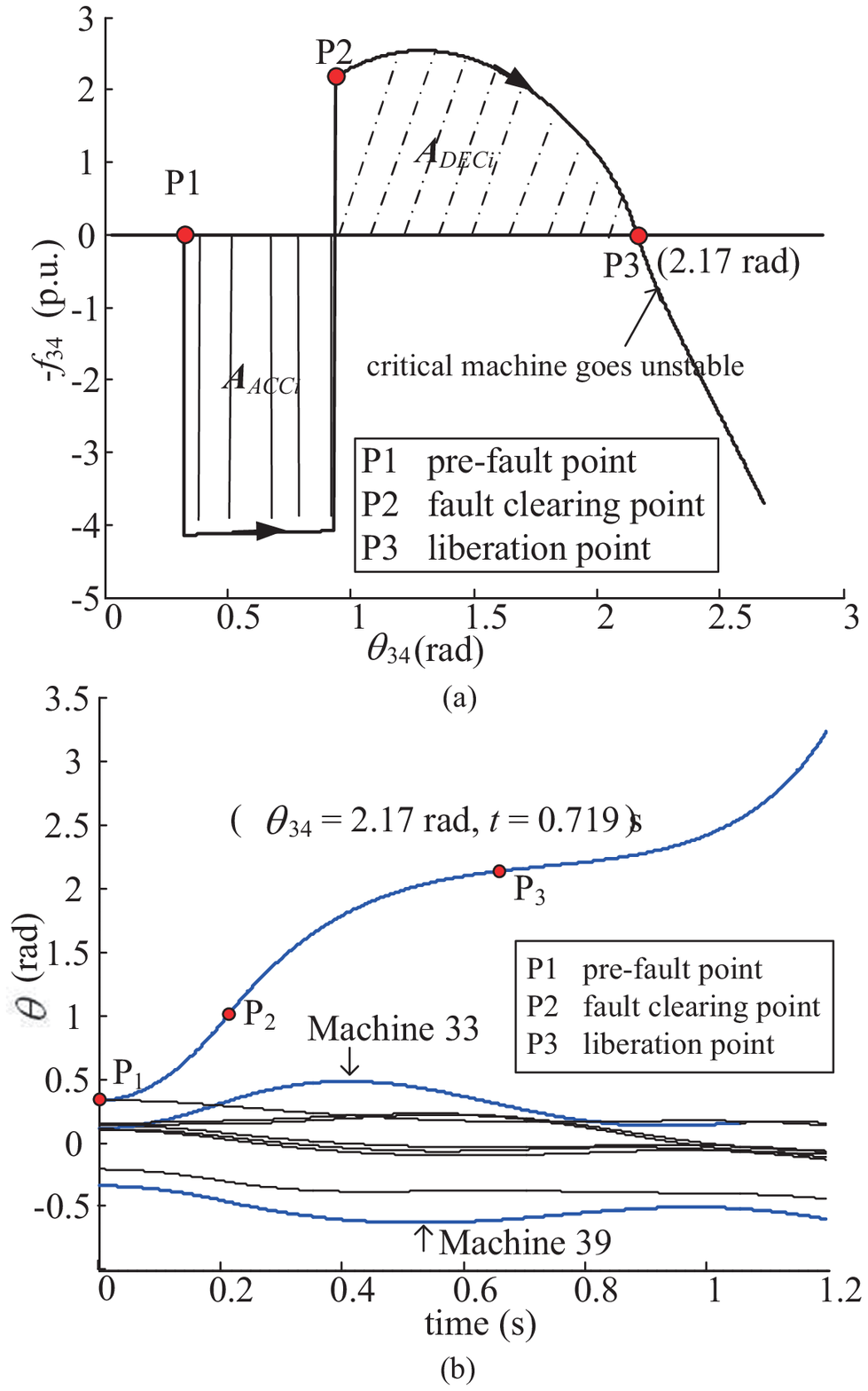}\\
  \setlength{\abovecaptionskip}{-5pt}
  \setlength{\belowcaptionskip}{0pt}
  \vspace{-2pt}
  \caption{ Simulations of the fault [TS-1, bus-34, 0.202s]. (a) Kimbark curve of Machine 34. (b) System trajectory}
  \label{Fig_Sim_Fal_4}
\end{figure}

Assume the Kimbark curve of the machine reaches the liberation point (P3) as in Fig. \ref{Fig_Sim_Fal_4}. Under this circumstance, both integral parts in (\ref{Eq_EAC_Pos_1}) can be seen as ``areas”, and the difference between the acceleration area and the deceleration area is the residual ``K.E.” of the machine at the liberation point (P3). Therefore, the machine can be judged as unstable as long as the acceleration area is larger than the deceleration area (i.e., the residual K.E. is positive at the liberation point), and the machine can be judged as stable if the acceleration area is equal to the deceleration area (i.e., the residual K.E. is strictly zero at the liberation point). \emph{This fully proves that EAC strictly holds for an individual machine.}

In power system transient stability analysis, critical machines are those a few severely disturbed machines, which are most possible to separate from the system, while non-critical machines are those slightly disturbed machines and they oscillate during post-fault period. Therefore, the transient behavior and corresponding Kimbark curve of a critical machine will be quite different from that of a non-critical machine, which will be analyzed in following sections. In this two-paper series the identification of the critical machines is fully based on the methodologies that were proposed in Ref \cite{Fouad1981Transient} and \cite{Haque1995Further} which consider critical machines as those severely disturbed machines with advanced angles \cite{Fouad1981Transient} and high acceleration ratios \cite{Haque1995Further}.

\section{Kimbark Curve of an Individual Machine}
\subsection{Kimbark Curve of an Unstable Critical Machin}

From numerous simulations, the representative Kimbark curve of a critical machine going unstable is shown in Fig. \ref{Fig_Sim_Fal_5} (a). The corresponding system trajectory is shown in Fig. \ref{Fig_Sim_Fal_5} (b). Machines 33, 34 and 39 are critical machines for the fault [TS-1, bus-34, 0.219s].

\begin{figure}
\vspace{5pt}
\captionsetup{name=Fig.,font={small},singlelinecheck=off,justification=raggedright}
  \includegraphics[width=3.5in,height=5.4in,keepaspectratio]{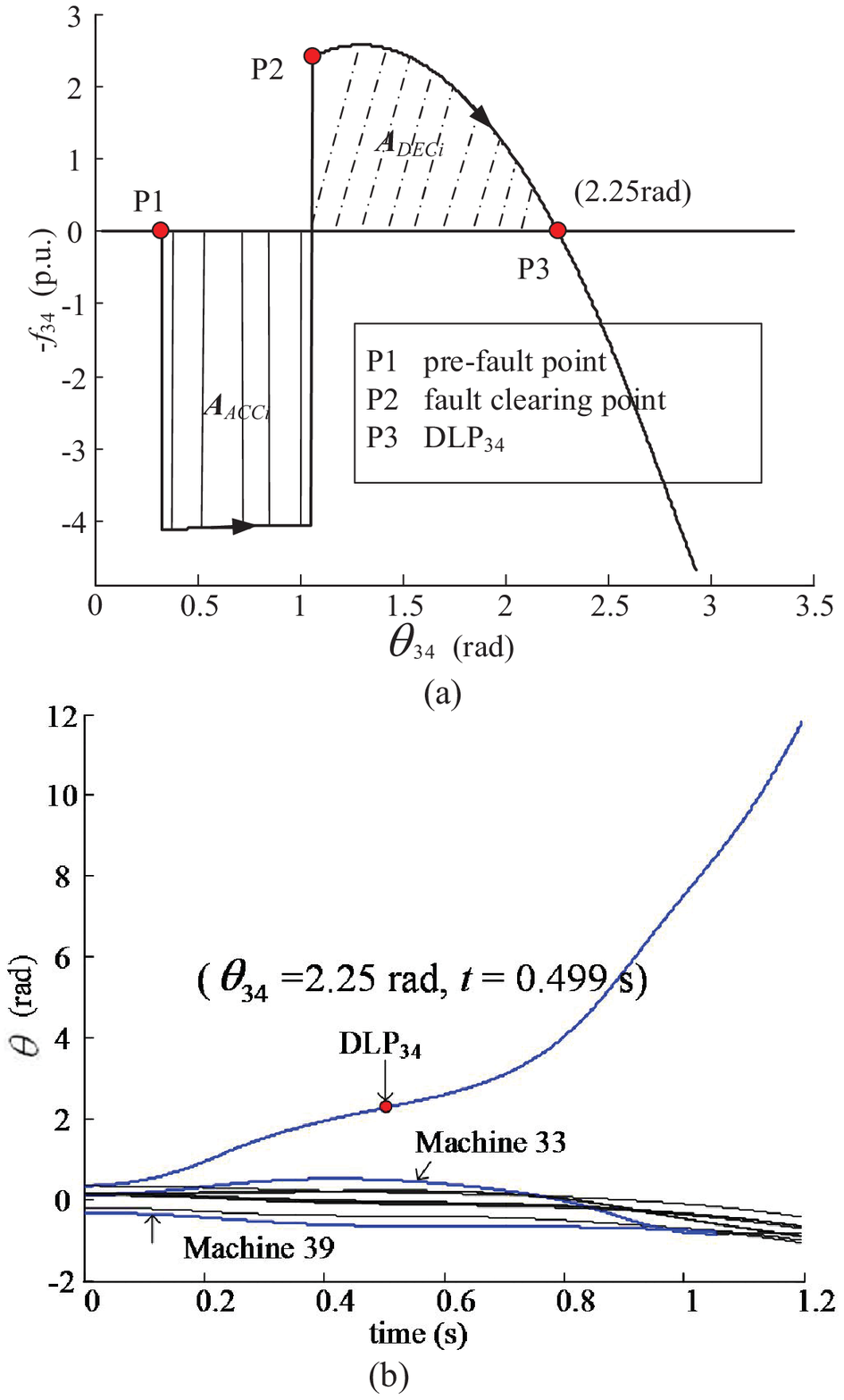}\\
  \setlength{\abovecaptionskip}{-5pt}
  \setlength{\belowcaptionskip}{0pt}
  \vspace{-2pt}
  \caption{Simulations of the fault [TS-1, bus-34, 0.219s]. (a) Kimbark curve of Machine 34. (b) System trajectory}
  \label{Fig_Sim_Fal_5}
\end{figure}

From Fig. \ref{Fig_Sim_Fal_5} (a), a critical machine firstly accelerates from P1 to P2 during fault-on period, then it decelerates from P2 to P3 after fault is cleared. Once system trajectory goes across P3, the critical machine will accelerate in COI reference and then separate from the system. Thus P3 can be defined as dynamic liberation point (DLP) of this unstable critical machine \cite{Stanton1989Analysis}.

Specifically, some critical machines may have negative velocity after fault clearing and finally anti-accelerates with time. In this case the Kimbark curve of the critical machine would seem to be “rotated”, as shown in Fig. \ref{Fig_Sim_Fal_6} (a).

\begin{figure}
\vspace{5pt}
\captionsetup{name=Fig.,font={small},singlelinecheck=off,justification=raggedright}
  \includegraphics[width=3.5in,height=5.4in,keepaspectratio]{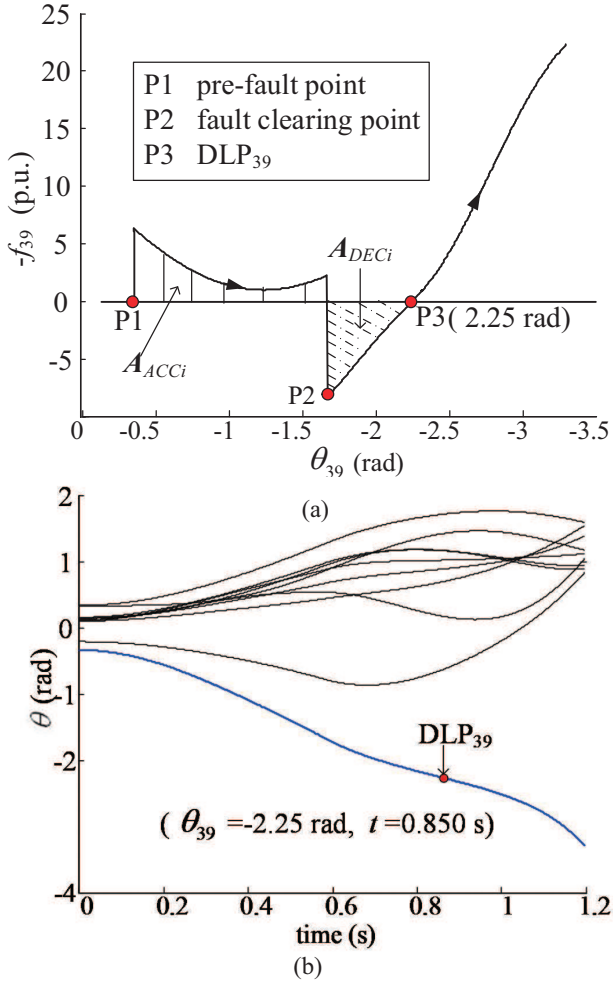}\\
  \setlength{\abovecaptionskip}{-5pt}
  \setlength{\belowcaptionskip}{0pt}
  \vspace{-2pt}
  \caption{Simulations of the fault [TS-1, bus-3, 0.580s]. (a) Kimbark curve of Machine 39. (b) System trajectory}
  \label{Fig_Sim_Fal_6}
\end{figure}

From Figs. \ref{Fig_Sim_Fal_4}-\ref{Fig_Sim_Fal_6}, the unstable case of the critical machine can be characterized by the occurrence of the DLP with $f_i$ of the machine being zero:

\begin{equation}\label{Eq_Uns_CMDLP}
\tilde{w_{i}}\neq{0}, f_{i}=0
  \setlength{\abovedisplayskip}{1pt}
  \setlength{\belowdisplayskip}{1pt}
\end{equation}

Eq. \ref{Eq_CSta_CMDLP} corrects the instability characterization of the critical machine in Ref. \cite{Stanton1989Analysis,Stanton1989Transient} because these two papers neglected the anti-accelerating case as shown in Fig. \ref{Fig_Sim_Fal_6} (a).

Following EAC, when couple machines go unstable, we have:

\begin{equation}\label{Eq_EAC_Uns_Mac}
A_{ACCij}>A_{DECij}
  \setlength{\abovedisplayskip}{1pt}
  \setlength{\belowdisplayskip}{1pt}
\end{equation}
where
\newline
$A_{ACCij}=\int_{\theta_{i}^{0}}^{\theta_{i}^{c}}(-f_{i}^{(F)})d\theta_{i}=\frac{1}{2}M_{i}\tilde{\omega_{i}^{(c)}}^{2}$
\newline
$A_{DECij}=\int_{\theta_{i}^{c}}^{\theta_{i}^{DLP}}(-f_{i}^{(F)})d\theta_{i}$

Notice that only the mismatch $f_{i}$ of machine $i$ is zero while mismatch of other machines are not zero at $\text{DLP}_{i}$. For more than one critical machines that go unstable after fault clearing, each critical machine corresponds to its unique DLP.

From analysis above, the Kimbark curve of an unstable critical machine has a clear ``accelerating-decelerating-accelerating” characteristic. Since the Kimbark curve is formulated from actual simulated system trajectory, the Kimbark curve of an unstable critical machine and its corresponding DLP varies with the change of the faults. For instance, $\text{DLP}_{34}$ in Fig. \ref{Fig_Sim_Fal_4} (a) is different from that in Fig. \ref{Fig_Sim_Fal_5} (a) if the system is subject to a different fault.

The Kimbark curve of an individual machine in the proposed method is a bit similar to that of the OMIB system as in the IEEAC method and SIME method. However, we emphasize that the EAC in the proposed method is strictly based on the Kimbark curve of an individual machine in COI reference, which is quite different from that in IEEAC method and SIME method that is based on the equivalent OMIB system in a multi-machine system.

\subsection{Kimbark Curve of a Stable Critical Machine }

From simulations, representative Kimbark curves of critical machines being stable are shown in Figs. \ref{Fig_Sim_Fal_7} (a-c). The system trajectory is shown in Fig. 7 (d). Machines 33, 34 and 39 are critical machines for the fault [TS-1, bus-34, 0.180s].

\begin{figure}
\vspace{5pt}
\captionsetup{name=Fig.,font={small},singlelinecheck=off,justification=raggedright}
  \includegraphics[width=3.6in,height=4.4in,keepaspectratio]{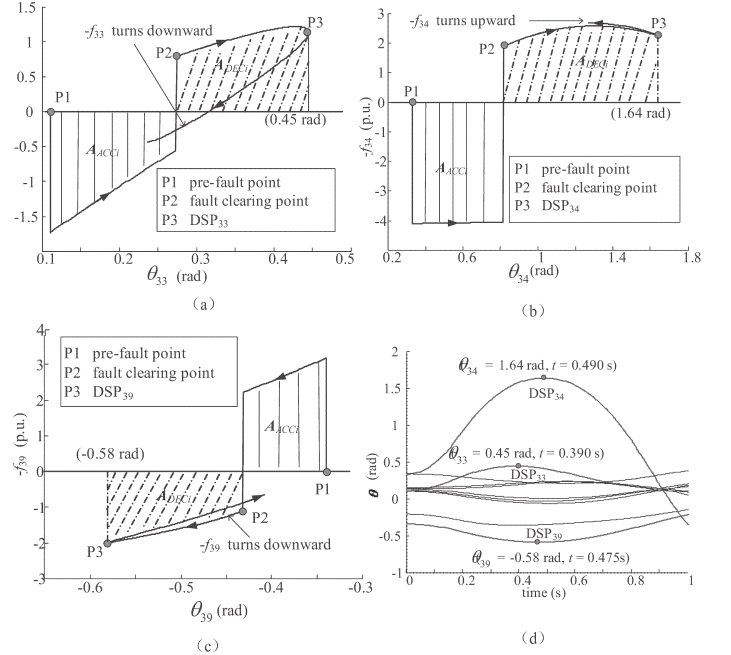}\\
  \setlength{\abovecaptionskip}{-5pt}
  \setlength{\belowcaptionskip}{0pt}
  \vspace{-2pt}
  \caption{ Simulations of the fault [TS-1, bus-34, 0.180s]. (a-c) Kimbark curves of Machines 33, 34 and 39. (d) System trajectory}
  \label{Fig_Sim_Fal_7}
\end{figure}

From Figs. \ref{Fig_Sim_Fal_7}(a-c), for a stable critical machine, the machine first accelerates from P1 to P2 during fault-on period, then it decelerates and the velocity of the machine then reaches zero at P3. Therefore, the machine never goes across DLP, i.e., $-f_i$ will not intersect with horizontal zero line. Instead, $-f_i$ might turn upward or turn downward because the oscillation of other machines may impede backtracking of the system trajectory. In this way the critical machine will not separate from the system and is maintained stable in the first swing, thus P3 with zero velocity and the maximum angle can be defined as the dynamic stationary point (DSP) of the critical machine.

From Fig. \ref{Fig_Sim_Fal_7}, the stable case of the critical machine can be characterized by the occurrence of the DSP with the velocity of the machine being zero:

\begin{equation}\label{Eq_Cri_Sta_DSP}
\tilde{w_{i}}={0}, f_{i}\neq{0}
  \setlength{\abovedisplayskip}{1pt}
  \setlength{\belowdisplayskip}{1pt}
\end{equation}

Eq. (\ref{Eq_Cri_Sta_DSP}) corrects the stability characterization of the critical machine in Ref. [10, 11] because these two papers neglected the anti-accelerating case as in Fig. \ref{Fig_Sim_Fal_7} (c).

Following IMEAC, when a critical machine is stable, we have:

\begin{equation}\label{Eq_EAC_Cri_Mac}
A_{ACCij}=A_{DECij}
  \setlength{\abovedisplayskip}{1pt}
  \setlength{\belowdisplayskip}{1pt}
\end{equation}
where
\newline
$A_{ACCij}=\int_{\theta_{i}^{c}}^{\theta_{i}^{DSP}}(-f_{i}^{(F)})d\theta_{i}$

In the Kimbark curve of the stable critical machine, DSP is the inflection point where $-f_i$ turns upward or downward. DSP describes first swing stability of a critical machine. At DSP of the stable critical machine $i$, only the velocity of the critical machine $i$ is zero while the velocity of other machines are nonzero. For the case that more than one critical machines are stable after fault clearing, each critical machine corresponds to its unique DSP. The Kimbark curve of a stable critical machine has a clear ``accelerating-decelerating” characteristic before DSP occurs, and the DSP of a stable critical machine will vary with the change of the faults.

\subsection{Kimbark Curve of a Critical-stable Critical Machine }

Following stable and unstable characterization of a critical machine, once the critical machine is critical stable, the machine is still ``stable” and $-f_i$ will inflect at the DSP of the critical stable machine (CDSP). However, the CDSP is special because it just falls in the zero horizontal line, which describes the critically stable state of the machine. Therefore, the critical stable case of a critical machine is characterized by:

\begin{equation}\label{Eq_Cri_StaC_DSP}
\tilde{w_{i}}={0}, f_{i}={0}
  \setlength{\abovedisplayskip}{1pt}
  \setlength{\belowdisplayskip}{1pt}
\end{equation}

The Kimbark curve of a critical stable machine and corresponding system trajectory are shown in Figs. \ref{Fig_Sim_Fal_8} (a) and (b), respectively. In this case Machine 34 is critical stable while Machines 33 and 39 are stable.

\begin{figure}
\vspace{5pt}
\captionsetup{name=Fig.,font={small},singlelinecheck=off,justification=raggedright}
  \includegraphics[width=3.5in,height=5.4in,keepaspectratio]{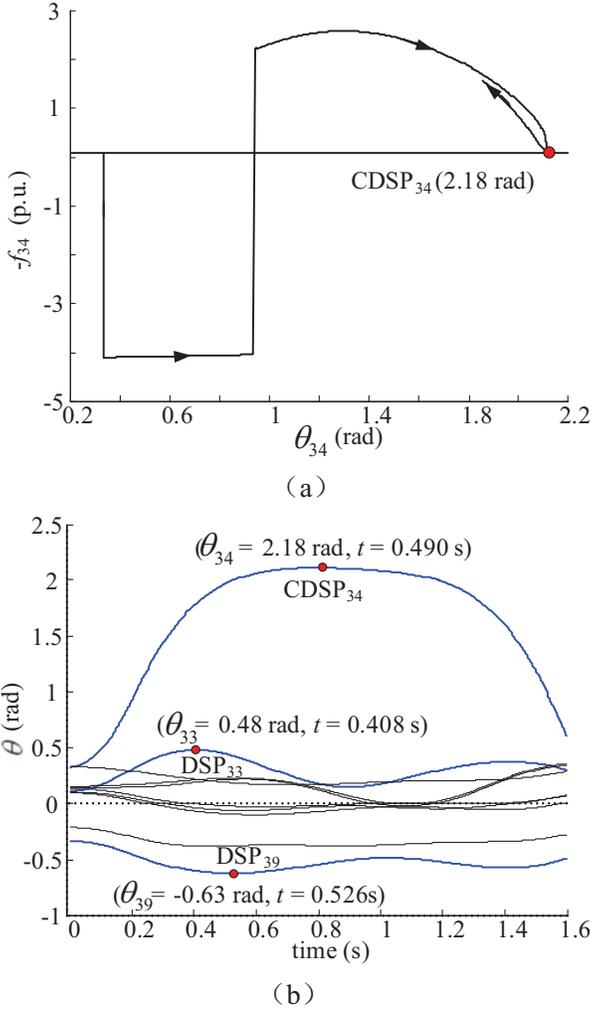}\\
  \setlength{\abovecaptionskip}{-5pt}
  \setlength{\belowcaptionskip}{0pt}
  \vspace{-2pt}
  \caption{Simulations of the fault [TS-1, bus-34, 0.201s]. (a) Kimbark curve of the Machine 34. (b) System trajectory}
  \label{Fig_Sim_Fal_8}
\end{figure}

\subsection{Kimbark Curve of a Non-critical Machine}

Since non-critical machines are majorities that are slightly disturbed by faults, non-critical machines generally maintain synchronism during post-fault period. In other words, the non-critical machines may oscillate with time, and the Kimbark curves of non-critical machines do not have a clear “accelerating-decelerating” characteristic compared with that of critical machines.

The distinctive feature of the Kimbark curve of the critical machine reveals the potential of using IMEAC when judging the stability of a critical machine. Therefore, the foremost crucial work for the multi-machine transient stability analysis is to depict the relationship between the stability of the system and that of critical machines, which will be analyzed in the following sections.

\section{Mapping between Trajectory and Kimbark Curve of an Individual Machine}

\subsection{Individual Machine Trajectory }

 Theoretically, the transient stability of the system should be explicitly expressed as the transient stability of the ``system trajectory”. If a system goes unstable, the separation of machines in the system would occur along time horizon. In this paper, the variation of the rotor angle of an individual machine along time horizon is defined as the “individual-machine trajectory” (IMT). A simulation case to demonstrate the system trajectory and IMTs is shown in Figs. \ref{Fig_Rot_Ang_9} and \ref{Fig_IMT_Ind_Mac_10}. Machines 37, 38 and 39 are critical machines in this case.

\begin{figure}
\vspace{5pt}
\captionsetup{name=Fig.,font={small},singlelinecheck=off,justification=raggedright}
  \includegraphics[width=3.5in,height=2.4in,keepaspectratio]{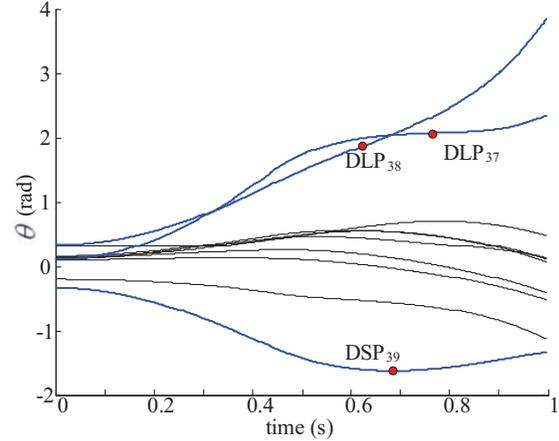}\\
  \setlength{\abovecaptionskip}{-5pt}
  \setlength{\belowcaptionskip}{0pt}
  \vspace{-2pt}
  \caption{Rotor angles of the system [TS-1, bus-2, 0.430s]}
  \label{Fig_Rot_Ang_9}
\end{figure}

\begin{figure}
\vspace{5pt}
\captionsetup{name=Fig.,font={small},singlelinecheck=off,justification=raggedright}
  \includegraphics[width=3.5in,height=4.8in,keepaspectratio]{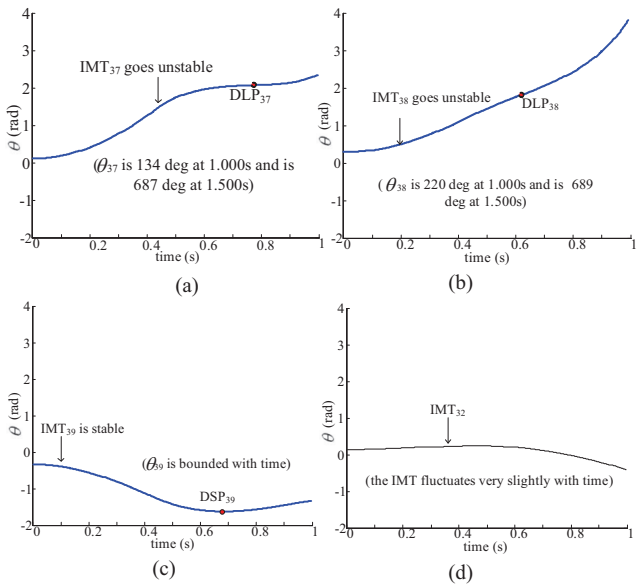}\\
  \setlength{\abovecaptionskip}{-5pt}
  \setlength{\belowcaptionskip}{0pt}
  \vspace{-2pt}
  \caption{ IMTs of individual machines [TS-1, bus-2, 0.430s]. (a-d) IMTs of Machines 37, 38, 39 and 32.}
  \label{Fig_IMT_Ind_Mac_10}
\end{figure}

The IMTs in Fig. \ref{Fig_IMT_Ind_Mac_10} reveal a commonly observed phenomenon in the power transient stability, i.e., after fault clearing the IMTs of critical machines fluctuate most severely, and they are most possible to separate from the system. Comparatively, IMTs of non-critical machines fluctuate slightly and these machines hardly separate from the system.
Depicting the variation of the IMT in a mathematical form, the IMT of a critical machine going unstable is identical to the rotor angle of the machine in COI reference going infinite with time, which can be expressed as:

\begin{equation}\label{Eq_IMT_Ins}
|\theta_{i,t}|=|\int_{t_{0}}^{t}\tilde{\omega_{i}}dt|=+\infty  \quad t=+\infty
  \setlength{\abovedisplayskip}{1pt}
  \setlength{\belowdisplayskip}{1pt}
\end{equation}

Comparatively, the IMT of a critical machine being stable is identical to the rotor angle of the machine in COI reference being bounded with time, which can be denoted as:
\begin{equation}\label{Eq_IMT_Sta}
|\theta_{i,t}|=|\int_{t_{0}}^{t}\tilde{\omega_{i}}dt|<\theta_{i}^{bound} \quad t\in{(0,+\infty)}
  \setlength{\abovedisplayskip}{1pt}
  \setlength{\belowdisplayskip}{1pt}
\end{equation}
where $\theta_{i}^{bound}$ is the upper bound of $\theta_{i,t}$.

Based on the analysis above, the original thinking of trajectory stability of the system can be merited as below:

(i) If IMTs of all critical machines are bounded, the separation of machines in the system is impossible to occur, and system can maintain stable (Fig \ref{Fig_IMT_Ind_Mac_10} (c)).

(ii) If IMTs of some critical machines go infinite along with time, the separation of machines in the system is certain to occur, and the system would go unstable (Figs \ref{Fig_IMT_Ind_Mac_10} (a,b)).

(iii) The IMTs of non-critical machines always fluctuate slightly and they hardly separate from the system, thus IMTs of non-critical machines are unable to cause system to go unstable (Fig \ref{Fig_IMT_Ind_Mac_10} (d)).

The statements above can be seen as the foundation of the proposed method. From the angle of the trajectory stability, the slight fluctuations of the IMTs of the non-critical machines have no effect to the instability of the system. Comparatively, those severely fluctuated IMTs of critical machines are most possible to go infinite and cause system to go unstable. Therefore, the following criterion is proposed for TSA:

{\it ``The system operator may only monitor stability of IMTs of critical machines during post-fault transient period.
Furthermore, the prime objective of the system operator is to find out the IMTs of unstable critical machines among all critical machines in the system, because only IMTs of unstable critical machines may cause system to go unstable.”
}

\subsection{3-dimensional Kimbark Curve of a Critical Machine}

In transient stability analysis, a significant defect of observing IMT is that the transient behavior of the machine is quite difficult to be depicted. To solve this problem, it is necessary to map the stability analysis of the IMT in the $t-\theta_{i}$ space to the $\theta_{i}-f_{i}$ space wherein IMEAC can be used to analyze the individual-machine stability.
In order to demonstrate the mapping between IMT and IMEAC, a 3-dimensional Kimbark curve (3DKC) of a critical machine in the $t-\theta_{i}-f_{i}$ space is proposed in this paper, as shown in Fig. \ref{Fig_3D_Kim_Cur_11}. All parameters in the 3DKC of the machine are fully formulated from the actual simulated system trajectory.

\begin{figure}
\vspace{5pt}
\captionsetup{name=Fig.,font={small},singlelinecheck=off,justification=raggedright}
  \includegraphics[width=3.5in,height=2.4in,keepaspectratio]{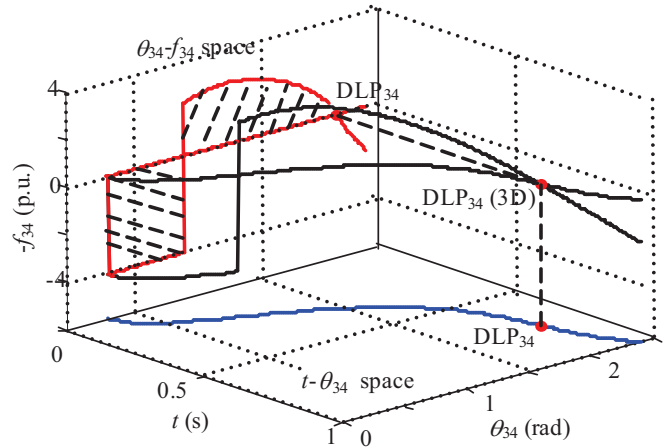}\\
  \setlength{\abovecaptionskip}{-5pt}
  \setlength{\belowcaptionskip}{0pt}
  \vspace{-2pt}
  \caption{3DKC of an unstable critical machine in a multi-machine system [TS-1, bus-34, 0.202s].}
  \label{Fig_3D_Kim_Cur_11}
\end{figure}

From Fig. \ref{Fig_3D_Kim_Cur_11}, by using the 3DKC of the critical machine, the IMT of the critical machine in the
$t-\theta_{i}$ space is mapped to the Kimbark curve of the machine in the $\theta_{i}-f_{i}$  space, and the stability of a critical machine can be easily measured by the occurrence of DLP or DSP in the Kimbark curve of this machine. This proves that the stability of IMT of a critical machine can be identified via IMEAC.

We extend the concept of the 3DKC of an individual machine to the stability evaluation of a system with n machines. Following the definition of IMT, it is obvious that the system trajectory can be seen as the ``set” that comprises of IMTs of all individual machines in the system. The mappings between system trajectory and 3DKCs of all individual machines in the system are shown in Fig. \ref{Fig_Map_3DKC_12}.

\begin{figure*}
\vspace{5pt}
\captionsetup{name=Fig.,font={small},singlelinecheck=off,justification=raggedright}
  \includegraphics[width=8in,height=5.4in,keepaspectratio]{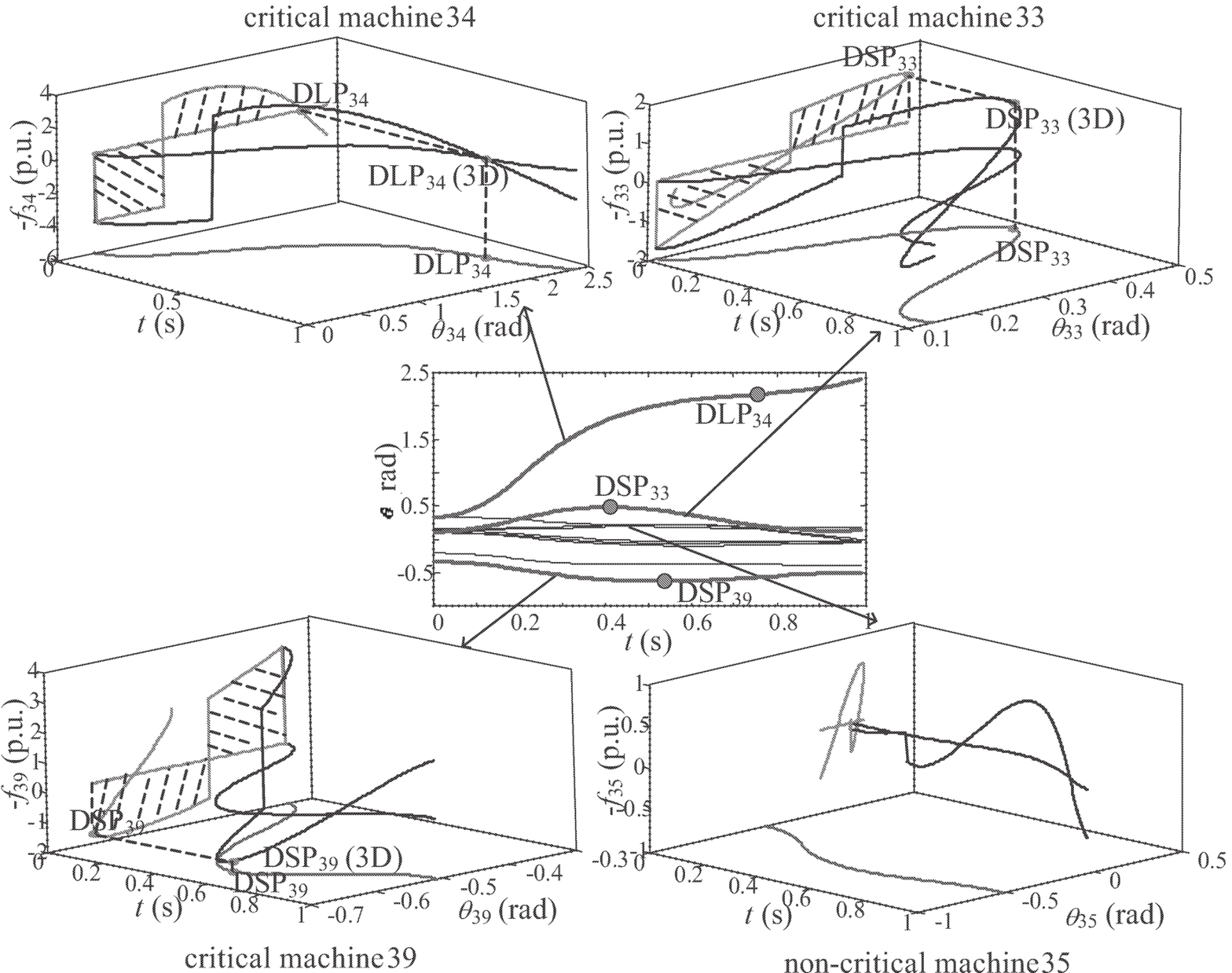}\\
  \setlength{\abovecaptionskip}{-5pt}
  \setlength{\belowcaptionskip}{0pt}
  \vspace{-2pt}
  \caption{ Mappings between system trajectory and 3DKCs of all individual machines in the system [TS-1, bus-34, 0.202s]}
  \label{Fig_Map_3DKC_12}
\end{figure*}

\section{Unity Principle}

\subsection{Only-one-machine Monitoring}

 From analysis in Section 4, since the instability of the system is determined by IMTs of unstable critical machines, the system operators can monitor the IMT of each critical machine in the system in parallel to identify the real unstable critical machine. Then one question emerges: could the instability of the system be evaluated if the system operator does not monitor all critical machines?

We extend the trajectory monitoring to an extreme “only-one-machine” way. Taking the case in Fig. \ref{Fig_IMT_Ind_Mac_10} for example, one can find that the IMTs of critical machines 37 and 38 both go unstable. Assume under an extreme circumstance that the system operator knows ten machines are operating in the system. However, he only focuses on Machine 37 and does not observe all the other machines in the system, as shown in Fig. 13. Under this extreme ``only-one-machine monitoring” circumstance, could the system still be defined as unstable?

Fig. \ref{Fig_IMT_Mon_13} intuitively demonstrates the mechanism of using IMEAC for TSA. From the figure, the system trajectory comprises of n IMTs and each machine’s IMT corresponds to its unique 3DKC, thus the system trajectory can be easily mapped into n 3DKCs. Among all 3DKCs, considering that only IMTs of critical machines may cause the instability of the system, the system operators only need to observe 3DKCs of critical machines by neglecting that of non-critical machines. Inside 3DKC of each critical machine, the stability of the machine is evaluated via IMEAC (i.e., the occurrence of DLP or DSP). Once one or more critical machines are found to go unstable, the system can be judged as unstable according to the trajectory stability theory.

\begin{figure}
\vspace{5pt}
\captionsetup{name=Fig.,font={small},singlelinecheck=off,justification=raggedright}
  \includegraphics[width=3.5in,height=2.4in,keepaspectratio]{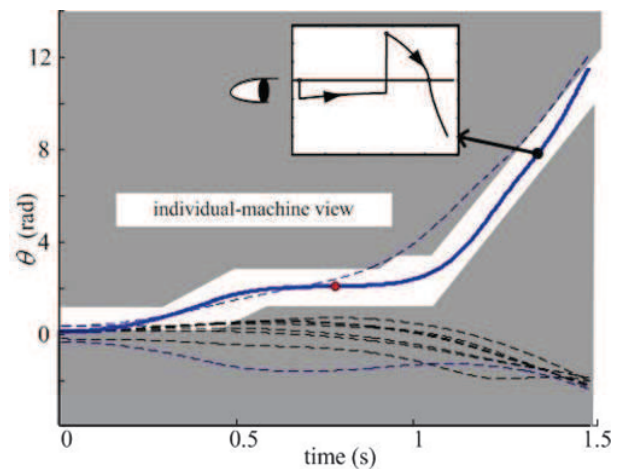}\\
  \setlength{\abovecaptionskip}{-5pt}
  \setlength{\belowcaptionskip}{0pt}
  \vspace{-2pt}
  \caption{Monitoring IMT of only one unstable machine [TS-1, bus-2, 0.430s]}
  \label{Fig_IMT_Mon_13}
\end{figure}

From Fig. \ref{Fig_IMT_Mon_13} one can see, in COI reference $\text{DLP}_{37}$ occurs at 0.777s and $\theta_{37}$ reaches 687 deg. at 1.500s. Under this extreme individual-machine monitoring circumstance, although IMTs of all the other machines in the system are not monitored, it is quite obvious that the system cannot be maintained stable because IMT37 keeps separating from the system with time.

\textbf{Theorem}: Transient instability of any one IMT determines the transient instability of the system trajectory.

\textbf{Proof}: Using Reductio-ad-absurdum, assume the system is stable when an IMT in the system already goes infinite with time. Following trajectory stability theory, this assumption is contradict to the sufficient and necessary condition such that the system should maintain stable, i.e., IMTs of all machines in the system should be bounded along time horizon (Section IV), thus theorem holds.

Following analysis above, the unity of individual-machine stability and system stability can be expressed as:

(I) \textit{The system can be judged as stable if all critical machines are stable.
}
(II)\textit{ The system can be judged as unstable as long as any one critical machine is found to go unstable.
}

Principles I and II substantially illustrates the unity of individual-machine stability and system stability. Especially, Principle II is of interest because it reflects that the transient stability of a multi-machine system may be monitored in an individual-machine way, and the instability of the system can be determined by any one unstable critical machine without monitoring all critical machines in the system. This provides a quite novel individual-machine angle for transient stability analysis in TSA.

\subsection{Occurrence of Multiple LOSPs of the System}

Following unity principle, since each unstable critical machine may cause system to go unstable, the DLP of each unstable critical machine can be seen as the LOSP of the system. Therefore, multiple LOSPs might exist along post-fault system trajectory if more than one critical machines goes unstable, and these LOSPs can be classified as below:

\textit{Leading LOSP}: The leading LOSP is defined as the first occurred DLP along time horizon;

\textit{Lagging LOSP}: The lagging LOSP is defined as the DLP that occurs later than the leading LOSP.

From definitions above, the system can be judged as unstable once the leading LOSP or the lagging LOSPs occur. However, the leading LOSP is certainly the most valuable for the system operators because the system starts separating at this point, as shown in Fig. \ref{Fig_Mul_LOSP_14}.

\begin{figure}
\vspace{5pt}
\captionsetup{name=Fig.,font={small},singlelinecheck=off,justification=raggedright}
  \includegraphics[width=3.5in,height=2.4in,keepaspectratio]{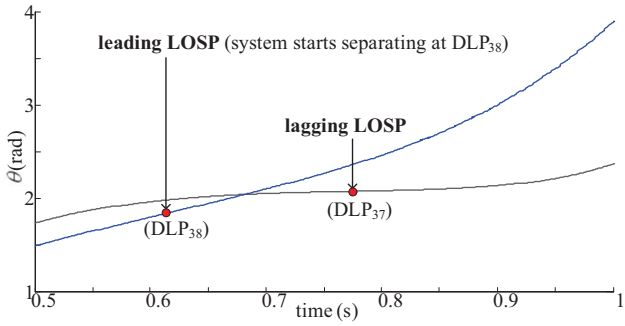}\\
  \setlength{\abovecaptionskip}{-5pt}
  \setlength{\belowcaptionskip}{0pt}
  \vspace{-2pt}
  \caption{Multiple LOSPs in a multimachine system [TS-1, bus-2, 0.430s]}
  \label{Fig_Mul_LOSP_14}
\end{figure}

\subsection{ Machine-by-machine Stability Judgement}

In actual TSA environment, the system operator may monitor each critical machine in parallel once fault is cleared, and the stability of a critical machine can be identified once the corresponding DLP or DSP occurs in the Kimbark curve. Yet, since DSPs and DLPs occur one after another along post-fault system trajectory as analyzed in Section II, the stability of critical machines in the system can only be identified in a ``machine-by-machine” way along time horizon. Furthermore, during this process the system can be judged as unstable immediately once the leading LOSP occurs without waiting for the stability judgements of the rest of critical machines. Detailed analysis will be provided in the next section.

\section{Case Studies}

\subsection{ Parallel Monitoring}

The case [TS-1, bus-2, 0.430s] is provided here to demonstrate the machine-by-machine stability judgement when using proposed method in TSA. The simulated system trajectory is shown in Fig. \ref{Fig_Par_Mon_15}.

\begin{figure*}
\vspace{5pt}
\captionsetup{name=Fig.,font={small},singlelinecheck=off,justification=raggedright}
  \includegraphics[width=6.5in,height=4.4in,keepaspectratio]{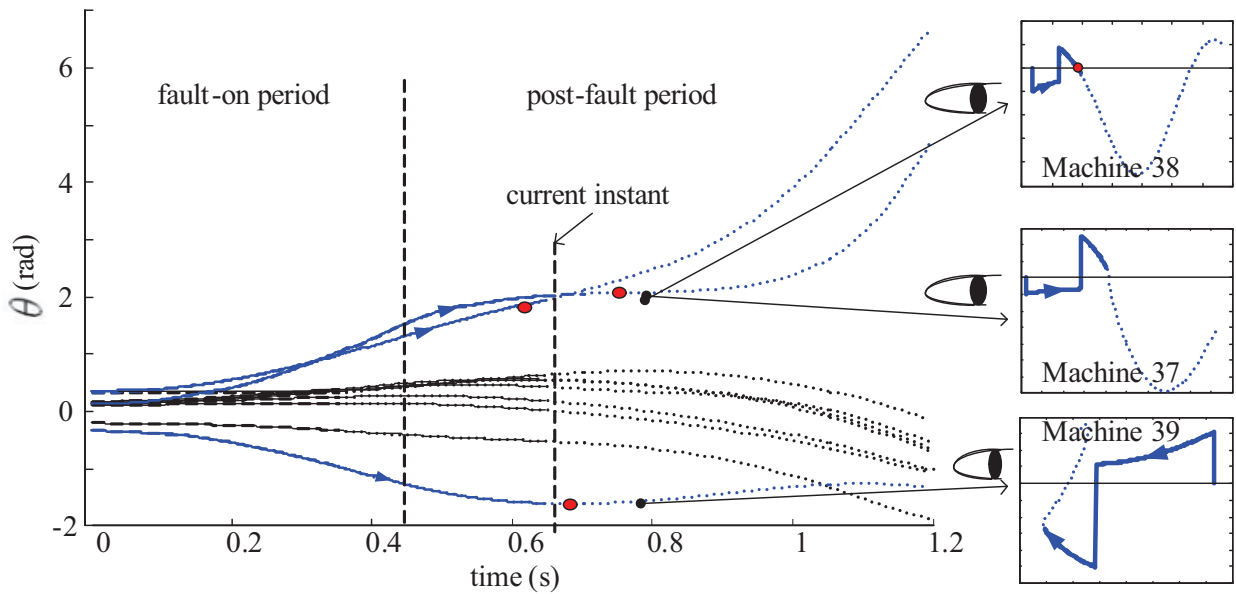}\\
  \setlength{\abovecaptionskip}{-5pt}
  \setlength{\belowcaptionskip}{0pt}
  \vspace{-2pt}
  \caption{Demonstration of parallel monitoring [TS-1, bus-2, 0.430s]}
  \label{Fig_Par_Mon_15}
\end{figure*}

In Fig. \ref{Fig_Par_Mon_15}, once fault is cleared, Machines 37, 38 and 39 are identified as critical machines. Therefore, the system operator monitors IMTs of these three critical machines in parallel by neglecting that of non-critical machines. Along time horizon the system operator can focus on following instants.

\textit{$DLP_{38}$ occurs (0.614s)}: Machine 38 is judged as unstable.

\textit{$DSP_{39}$ occurs (0.686s)}: Machine 39 is judged as stable.

\textit{$DLP_{37}$ occurs (0.777s)}: Machine 37 is judged as unstable.

The stability of the system is judged as below:

\textit{$DLP_{38}$ occurs (0.614s)}: $\text{DLP}_{38}$ is defined as leading LOSP, and the system is judged as unstable.

\textit{$DLP_{37}$ occurs (0.777s)}: $\text{DLP}_{37}$ is defined as lagging LOSP, yet the system has already gone unstable for a while.

From analysis above, $\text{DLP}_{38}$ and $\text{DLP}_{37}$ are the leading LOSP and lagging LOSP, respectively. $f_{i}s$ of all machines in the system at the instant of the occurrence of $\text{DLP}_{37}$ (0.777s) are shown in Table \ref{Tab_Acc_Mac}.

\begin{table}
\captionsetup{name=\textsc{Table}}
\vspace{5pt}
\centering
\setlength{\belowcaptionskip}{0pt}
\caption{\textsc{Acceleration power of machines at} $\text{DLP}_37$}
\vspace{-4pt}
\setlength{\abovecaptionskip}{2pt}
\begin{tabular}{cccc}
\hline
Generator  & $f_{i} (p.u.)$ & Generator & $f_{i} (p.u.)$ \\
\hline
37 & \underline{0.00}    & 32             & 0.77          \\
38 &  -2.48             & 33             & 0.04          \\
39 &  -3.59             & 34             & 1.14          \\
30 &  1.53              & 35             & 1.31          \\
31 &  0.55              & 36             & 0.70          \\
\hline
\end{tabular}
\label{Tab_Acc_Mac}
\end{table}

From Table \ref{Tab_Acc_Mac} one can see, at 0.777s only $f_{37}$ is zero while $f_i$ of other machines are not zero, thus $\text{DLP}_37$ is only the acceleration point of Machine 37, which is meaningless to the stability analysis of other critical machines in the system.

\subsection{Only-one-machine Monitoring}

For the case given in Fig. \ref{Fig_Par_Mon_15}, assume that the system operator monitors only one critical machine and misses monitoring the other two critical machines. Under such circumstance, three different cases are shown as below:

\textit{a) Only Machine 38 is monitored}

In this case the system can be judged as unstable by the only monitored unstable Machine 38 via unity principle, and the leading LOSP can be obtained.

\textit{b) Only Machine 37 is monitored}

In this case the system can also be judged as unstable by the only monitored unstable Machine 37. Yet, the leading LOSP cannot be obtained. Only lagging LOSP can be obtained.

\textit{c) Only Machine 39 is monitored}

In this case the system operator cannot confirm that the system is stable or not by the only-monitored stable Machine 39, and the rest of critical machines still need to be monitored to confirm the instability of the system.
The individual-machine monitoring cases above indicate that each critical machine’s “status” for transient stability analysis in the system is different, which will be analyzed in the companion paper.

\section{Conclusion and discussion}

Through the analysis of this paper, one can conclude following:

(i) EAC is proved to strictly hold for a critical machine. The Kimbark curve of a critical machine exhibits a strong “accelerating-decelerating” characteristic.

(ii) The critical machines are most possible to separate from the system, and the system operator may only focus on analyzing the stability of critical machines.

(iii) A critical machine going unstable in $\theta_{i}-f_{i}$ space is identical to the IMT of the critical machine going unstable in $t-\theta_{i}$ space.

(iv) The unity principle indicates that monitoring the transient stability of the multi-machine system can be treated from an individual-machine angle, and transient instability of the multi-machine system can be determined by any one unstable critical machine.

In the companion paper, the application of the IMEAC and individual-machine stability judgement will be analyzed, which may demonstrate the effectiveness of using proposed method in TSA.

\end{document}